\documentclass{emulateapj}

\shorttitle{Thermal Radio Emission from HB3?}

\shortauthors{Uro{\v s}evi{\'c}, Pannuti \& Leahy}

\begin{document}

\title{An Analysis of the Broadband (22-3900 MHz) Radio Spectrum of HB3
(G132.7$+$1.3): The Detection of Thermal Radio Emission from an
Evolved Supernova Remnant?}

\author{D. Uro{\v s}evi{\'c}\altaffilmark{1,2}, T. G. Pannuti
\altaffilmark{3,4} and D. Leahy\altaffilmark{5}}

\altaffiltext{1}{Department of Astronomy, Faculty of Mathematics,
University of Belgrade, Studentski trg 16, 11000 Belgrade, Serbia
and Montenegro; dejanu@matf.bg.ac.yu}

\altaffiltext{2}{Isaac Newton Institute of Chile, Yugoslavia Branch}

\altaffiltext{3}{Space Science Center, Morehead State University,
200A Chandler Place, Morehead, KY 40351 USA;
t.pannuti@morehead-st.edu}


\altaffiltext{4}{Guest User, Canadian Astronomy Data Center, which
is operated by the Dominion Astrophysical Observatory for the
National Research Council of Canada's Herzberg Institute for
Astrophysics.}

\altaffiltext{5}{Department of Physics and Astronomy, University
of Calgary, Calgary, AB, Canada T2N 1N4; leahy@ucalgary.ca}

\keywords{radiation mechanisms: thermal--- radio continuum:
general --- supernova remnants --- ISM: individual:
HB3(G132.7+1.3)}

\begin{abstract}
We present an analysis of the broadband radio spectrum (from 22 to
3900 MHz) of the Galactic supernova remnant (SNR) HB3
(G132.7$+$1.3). Published observations have revealed that a
curvature is present in the radio spectrum of this SNR, indicating
that a single synchrotron component appears is insufficient to
adequately fit the spectrum. We present here a fit to this
spectrum using a combination of a synchrotron component and a
thermal bremsstrahlung component. We discuss properties of this
latter component and estimate the ambient density implied by the
presence of this component to be $n$ $\sim$ 10 cm$^{-3}$. We have
also analyzed extracted X-ray spectra from archived {\it ASCA} GIS
observations of different regions of HB3 to obtain independent
estimates of the density of the surrounding interstellar medium
(ISM). From this analysis, we have derived electron densities of
0.1-0.4 $f$$^{-1/2}$ cm$^{-3}$ for the ISM for the three different
regions of the SNR, where $f$ is the volume filling factor. By
comparing these density estimates with the estimate derived from
the thermal bremsstrahlung component, we argue that the radio
thermal bremsstrahlung emission is emitted from a thin shell
enclosing HB3. The presence of this thermal bremsstrahlung
component in the radio spectrum of HB3 suggests that this SNR is
in fact interacting with an adjacent molecular cloud associated
with the HII region W3. By extension, we argue that the presence
of thermal emission at radio wavelengths may be a useful tool for
identifying interactions between SNRs and molecular clouds, and for estimating the ambient density
near SNRs using radio continuum data.
\end{abstract}

\section{Introduction}

It has been firmly established that the dominant emission
mechanism from Galactic supernova remnants (SNRs) at radio
frequencies is synchrotron emission. This conclusion has been
reached based on the measured spectral index $\alpha$ of the
observed radio emission ($S$$_{\nu}$ $\sim$ $\nu$$^{-\alpha}$)from
these sources ranges from $\approx$ 0.3 - 0.7 \citep{Green06} and
-- in some cases -- the detection of polarized radio emission.
Typically, the radio emission from Galactic SNRs can be modeled by
a single power law: however, in the cases of some SNRs,
observations over a very broad range of radio frequencies reveal a
curvature in the spectra of these sources. Two scenarios have been
presented in the literature to explain this observed curvature:
the first scenario invokes spectral index variations due to
different populations of synchrotron emitting electrons associated
with the SNR (e.g., \citet{Tian05}), while the second scenario
argues for the presence of a thermal bremsstrahlung component in
the radio emission from the SNR in addition to the synchrotron
component. This thermal bremsstrahlung component is expected to be
most prominent for evolved SNRs interacting with molecular clouds
\citep{UrosevicDP03a, UrosevicDP03b, UrosevicP05}.
\par
To critically evaluate these two proposed scenarios, we present an
analysis of the broadband (38-3900 MHz) radio spectrum of one
Galactic SNR known to feature a curvature in its radio spectrum,
HB3 (G132.7$+$1.3). This source has been the subject of extensive
radio continuum, HI and OH observations \citep{HBH53, Landecker87,
Routledge91, Koralesky98, Tian05} which have revealed a shell-like
radio morphology for this source with an angular diameter of
approximately 80 arcminutes. The spectral index of this source is
$\alpha$ = 0.4 and the measured integrated flux density at 1 GHz
is 45 Jy \citep{Green06}: radio observations of this SNR are
complicated by confusing thermal emission from the adjacent HII
region W3. Thermal emission from an X-ray-emitting plasma located
in the interior of HB3 was discovered by {\it Einstein} and
described by \citet{Leahy85}. This X-ray emission is seen to lie
entirely within the radio shell of HB3 \citep{Leahy85}: this
combination of a radio shell morphology with a center-filled
thermal X-ray morphology has led to the classification of HB3 as a
mixed-morphology SNR \citep{Rho98}. In this paper, we separately
model the broadband radio spectrum of HB3 with a thermal
bremsstrahlung component combined with a synchrotron component:
from this modelling we obtain an estimate of the density of the
ambient interstellar medium (ISM) surrounding HB3. We compare this
estimate to ambient density estimates obtained from analyzing
extracted X-ray spectra of this SNR: we then discuss a possible
interaction between HB3 and W3.

\section{Modeling Thermal Bremsstrahlung Emission at Radio Frequencies from
HB3\label{ThermalBremSection}}

As described in previous works
\citep{UrosevicDP03a,UrosevicDP03b,UrosevicP05}, thermal
bremsstrahlung emission may become a significant emission process
at radio frequencies for evolved SNRs which are expanding into
regions of the interstellar medium (ISM) with enhanced densities
such as a molecular cloud. The volume emissivity
$\varepsilon$$_{\nu}$ (in CGS units) of thermal bremsstrahlung
emission for an ionized gas cloud is proportional to the square of
the electron (or ion) volume density $n$, i.e.,

\begin{equation}
\varepsilon_\nu [\mbox{ergs~s$^{-1}$~cm$^{-3}$~Hz$^{-1}$}] = 7
\times 10^{-38}{n^2 T^{-1/2}} \label{SigmaBremsEqn}
\end{equation}

\noindent \citep{Rohlfs96}, where $T$ is the thermodynamic
temperature of the ISM. The presence of a significant amount of
thermal bremsstrahlung emission will produce a curvature in the
observed radio spectrum of the SNR, particularly for frequencies
of 1 GHz and greater. \citet{Tian05} showed that spectral
flattening exists in the radio spectrum of HB3 at higher
frequencies and argued that this flattening originates from
synchrotron emission from different populations of electrons with
a range of energy (spectral) indices in low, middle and high radio
frequency domains. For comparison, we have modelled the radio
spectrum of HB3 using a single synchrotron component combined with
an additional thermal bremsstrahlung component. For this
artificially generated spectrum, we assume that the amount of
thermal bremsstrahlung emission is 40\% and the amount of
synchrotron emission is 60\% of total emission at 1GHz. In
addition, we generated an artificial spectrum using an
approximately ``pure" synchrotron component with a spectral index
$\alpha$ = 0.65. This value for $\alpha$ was obtained by fitting
the radio flux density from HB3 over the low frequency range 22 to
178 MHz (we expect that the thermal component should account for
10--25\% of the observed radio flux density over this frequency
range) and a thermal component with a spectral index $\alpha$ =
0.1. Based on these parameters, we calculated the frequency
spectrum over 10$^7$ -- 10$^{11}$ Hz. In Figure 1 (left) we
present both the resultant spectrum (as plotted with a thick line)
and the measured flux densities for HB3 over this frequency range
as compiled by \citet{Tian05} and references therein. From
inspection of this Figure it is clear that the inclusion of a
thermal component produces a noticeable bend in the radio spectrum
which fits the observed radio spectrum well. We therefore argue
that the total radio spectrum of HB3 is successfully modelled as
the sum of synchrotron and thermal bremsstrahlung components.
\par
Based on the properties of the thermal component used to fit the
radio spectrum of HB3, we can estimate the ambient density of the
ISM into which the SNR is expanding as follows. Using values
obtained from radio observations (such as flux density
$S$$_{\nu}$, diameter $D$ and thickness of SNR shell $s$) and
assuming a distance to HB3 of 2 kpc \citep{Routledge91} and an
electron temperature $T$ = 10$^4$ K, we use the relation given in
Equation 1 to calculate the volume emissivity. Specifically,
\citet{Tian05} reported $S$$_{\rm{1~GHz}}$ = 50 Jy, $D$ = 70 pc.
Using $s$ = 0.05$D$ as a reasonable estimate for shell thickness
of evolved SNRs, we obtain $\varepsilon$$_{\nu}$ =
1.67$\times$10$^{-37}$ ergs sec$^{-1}$ cm$^{-3}$ Hz$^{-1}$. From
Equation \ref{SigmaBremsEqn} and assuming that the thermal
component produces 40\% of the total radio emission from HB3 at 1
GHz, we estimate that the density of the ISM into which HB3 is
expanding is $\approx$ 10 cm$^{-3}$.


\section{X-ray Emission from HB3\label{XraySection}}

Because the X-ray emission from HB3 extends over such a large
angular extent on the sky, pointed observations were conducted
with the Advanced Satellite for Cosmology and Astrophysics ({\it
ASCA}) of three different regions (hereafter referred to as the
northern region, the central region and the southern region) to
ensure that virtually all of the X-ray emission from HB3 was
sampled. The bulk of this emission originates from the central
region where a prominent ring (approximately 35 arcminutes in
diameter) of emission is seen \citep{Leahy85}. The significant
variation in the X-ray brightness of HB3 from one portion of the
SNR to another clearly indicates that a wide range exists in the
density of the X-ray emitting gas. To help quantify the range of
this variation, we analyzed extracted spectra of the emission from
each of the three sampled regions as sampled by the two Gas
Imaging Spectrometers (GIS) -- denoted as GIS2 and GIS3 -- which
were aboard {\it ASCA} and collected data during these
observations. Details of the GIS observations of HB3 are provided
in Table \ref{Table1}.
\par
The data reduction process for the two observations were conducted
using the ``XSELECT" program (Version 2.2), which is available
from the High Energy Astrophysics Science Archive Research Center
(HEASARC \footnote{HEASARC is a service of the Laboratory for High
Energy Astrophysics (LHEA) at the National Aeronautics and Space
Administration Goddard Space Flight Center (NASA/GSFC) and the
High Energy Astrophysics Division of the Smithsonian Astrophysical
Observatory (SAO). For more information on HEASARC, please see
http://heasarc.gsfc.nasa.gov.}). We extracted both GIS2 and GIS3
spectra using the following extraction regions: for the northern
region, we used a circular region with a radius of
$\sim$10.3$\arcmin$ (centered at RA (J2000.0) 02$^h$ 20$^m$
06$^s$, Dec (J2000.0) $+$63$^{\circ}$ 17$\arcmin$ 44$\arcsec$);
for the central region, we used an elliptical region with radii
14.7$\arcmin$ $\times$ 14.2$\arcmin$ (centered at RA (J2000.0)
02$^h$ 17$^m$ 17$^s$, Dec (J2000.0) $+$62$^{\circ}$ 41$\arcmin$
49$\arcsec$) and finally for the southern region we used a
circular region with a radius of $\sim$12.3$\arcmin$ (centered at
RA (J2000.0) 02$^h$ 18$^m$ 15.3$^s$, Dec (J2000.0) $+$62$^{\circ}$
10$\arcmin$ 05$\arcsec$). The spectra were extracted in the format
of pulse-height amplitude (PHA) files as well as images and event
files: high bit-rate and medium bit-rate telemetry data were used
and the standard REV2 screening criteria were applied. For the
analysis of the GIS2 and GIS3 spectra, we used the standard
response matrix files (RMFs) and the corresponding ancillary
response files (ARFs) were prepared using the program ``ASCAARF"
(Version 3.10) from the FTOOLS software
package\footnote{``ASCAARF" and all of the other programs
mentioned in this Section are part of the FTOOLS software package.
For more information on this software package, please see
http://heasarc.gsfc.nasa.gov/ftools/.} \citep{Blackburn95}.
Background spectra and images for both GIS2 and GIS3 were prepared
using the FTOOL ``MKGISBGD" (Version 1.6). In Figure 1 (right), we
present the total band (0.7-10.0 keV) mosaicked {\it ASCA}
GIS2+GIS3 image of X-ray emission from HB3 with radio (1420 MHz)
contours overlaid.
\par
To obtain an independent estimate of the ambient density of the
ISM surrounding HB3, we performed spectral fitting on the
extracted {\it ASCA} GIS spectra for the three different regions
of this SNR. For each of the three regions, we simultaneously fit
the extracted GIS2 and GIS3 spectra with a collisional plasma
model component with solar elemental abundances known as APEC
\citep{Smith01}\footnote{Also see
http://hea-www.haravrd.edu/APEC.} combined with the PHABS
component to model interstellar extinction. Using this combination
of components, we were able to derive statistically acceptable
fits ($\chi$$^2$$_{\nu}$ $\sim$ 1.2 or less) for the extracted
spectra of the northern and southern regions, with column
densities ranging from $N$$_H$ $\approx$ 0.2 -- 0.6 $\times$
10$^{22}$ cm$^{-2}$ and temperatures $\it{kT}$ $\approx$ 0.4 --
0.6 keV. A statistically acceptable fit to the extracted GIS
spectra of the central region could not be obtained with this
combination of components. Prominent emission lines associated
with magnesium and silicon were seen in the extracted spectra, so
a collisional plasma model component with variable elemental
abundances -- VAPEC -- was implemented with the abundances of
those two elements allowed to vary. A Gaussian component was also
added to account for an emission line of unknown origin seen near
1.2 keV. A statistically acceptable fit still was not obtained, so
additional attempts to fit the spectra using a second component --
either an additional APEC component with a higher temperature or a
power law component -- were made. A statistically acceptable fit
was at last obtained after the addition of either of these later
two components. The column densities and temperatures of the lower
temperature thermal components are comparable in the two fits
($N$$_H$ $\sim$ 0.4-0.5$\times$10$^{22}$ cm$^{-2}$ and ${\it kT}$
$\sim$ 0.2 keV, respectively). The temperature of the higher
temperature component is ${\it kT}$ $\approx$ 2.2 keV while the
photon index of the power law component is $\Gamma$ $\approx$ 2.6.
We note that \citet{Leahy85} has already discussed the presence of
two thermal components in the X-ray emission from the central
region of HB3 as revealed by {\it Einstein} observations. We also
note that a gradient is seen in the column densities of the three
different regions, decreasing in proceeding from south to north:
this result is consistent with the presence of denser material
seen in projection toward the southern half of HB3. We summarize
the results of these spectral fits in Table \ref{Table2}. A
separate analysis of the ASCA data for HB3 has also been analyzed
and discussed by \citet{LS06}.
\par
Using the value for the emission measure of the soft thermal
component as derived in fits to the extracted GIS spectra for all
three regions, we calculated the corresponding electron density of
the ambient ISM for each region. We set each emission measure
equal to (10$^{-14}$/4$\pi$$d$$^2$)$\times$$n$$_e$
$\times$$n$$_H$$\times$$f$$\times$$V$, where $d$ is the distance
to HB3 (assumed to be 2 kpc), $n$$_e$ and $n$$_H$ are the electron
and hydrogen densities (with $n$$_H$ = 1.2 $n$$_e$), $f$ is the
volume filling factor and finally $V$ is the volume (assuming that
the depth of the emitting region is equal to its radius). From
this relation, we calculate electron densities $n$$_e$ $\approx$
0.4 $f$$^{-1/2}$ cm$^{-3}$ for the central region and $n$$_e$
$\approx$ 0.1 $f$$^{-1/2}$ cm$^{-3}$ for the northern and southern
regions (see Table \ref{Table2}).

\bigskip
\bigskip
\bigskip

\section{Discussion\label{DiscussSection}}

For comparison to Tian and Leahy (2005): Table 2 there summarizes
the spectral variations: the spectral indices range from 0 to 0.9
with most values between 0.3 and 0.7. The boxes nearest W3 are 9
and 13: they both contain flatter indices than 0.3, especially box
13. The variable spectral index was suggested to be due to
multiple populations of electrons since the normal case of a
single population with steeper index at higher energies does not
work to explain the radio spectrum (e.g. their Fig. 3). We are
suggesting here that a more natural explanation is synchrotron
emission which dominates at lower frequencies and bremsstrahlung
emission which dominates at higher frequencies. This is supported
by the high thermal electron densities derived from the X-ray
observations. One can still produce spatial variation of observed
spectral index caused by the spatial variation of relativistic
electrons and magnetic field (for the synchrotron component) and
of thermal electrons (for the bremsstrahlung component). Here we
present only a single fit to illustrate that the general idea of
synchrotron plus bremsstrahlung emission is viable.
\par
By comparing the densities implied by our fit to the broadband
radio spectrum with a thermal bremsstrahlung component with the
densities implied by our X-ray observations, we argue that the
density profile is such that ambient material is gathered in a
thin shell on the outer edge of HB3: in fact, the observed thermal
bremsstrahlung emission at radio frequencies is produced from this
shell. We also claim that the distribution of this material in a
shell indicates that HB3 is in fact interacting with the adjacent
molecular cloud W3. There has been some debate previously in the
literature regarding such an interaction: \citet{Routledge91}
described the detection (through CO observations) of the molecular
bar feature that is partially surrounded by continuum emission
from HB3, suggesting that HB3 is indeed interacting with this
cloud (and, by extension, W3). However, \citet{Koralesky98} failed
to find shock-excited maser emission at the nominal boundary
between the HII region and the SNR that could clearly be
associated with the shock of HB3. Our results supports interaction
between HB3 and W3; we also note that the gradient in the column
densities inferred by the X-ray observations is consistent with
the geometry of the known cloud and further supports such an
interaction. In fact, the detection of thermal emission at radio
frequencies from Galactic SNRs may be a crucial new tool in
determining whether these sources are interacting with adjacent
molecular clouds and for estimating the ambient density near SNRs
using radio continuum data.

\acknowledgments We thank an anonymous referee for valuable
comments and N. Duric for helpful discussions. D.U. would like to
thank T. Angelov and B. Arbutina for discussion on thermal
emission from SNRs at radio frequencies. T. G. P. would like to
thank J. Rho for discussions of the properties of mixed-morphology
SNRs. This work is part of the Project No. 146012 supported by the
Ministry of Science and Environmental Protection of Serbia.




\begin{figure}
\centerline{\epsscale{0.8}
\includegraphics[width=0.4\linewidth]{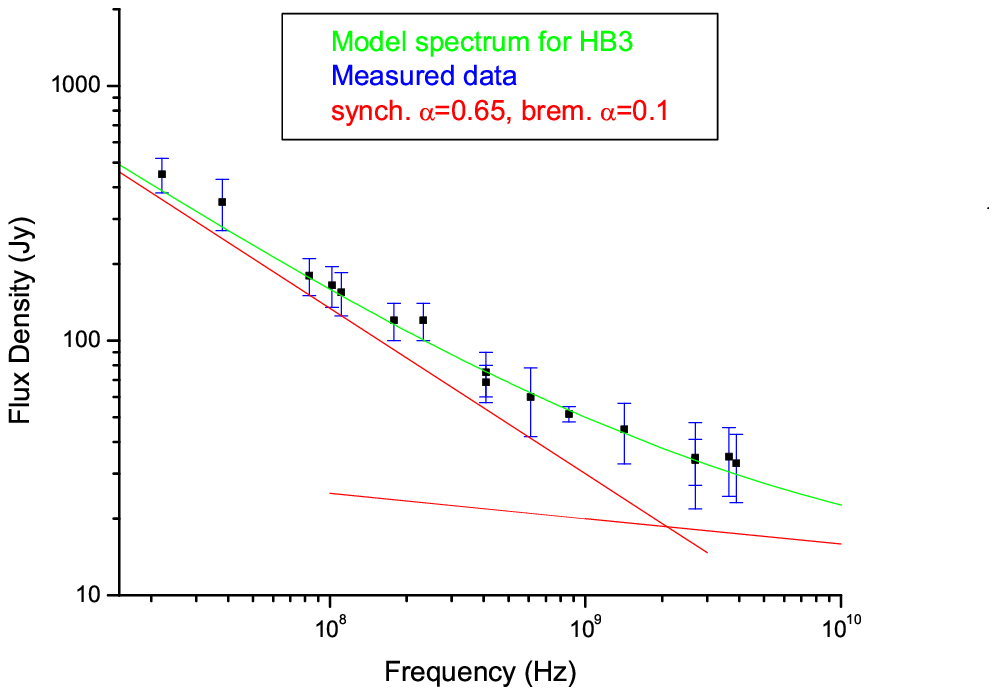}
 \includegraphics[width=0.3\linewidth]{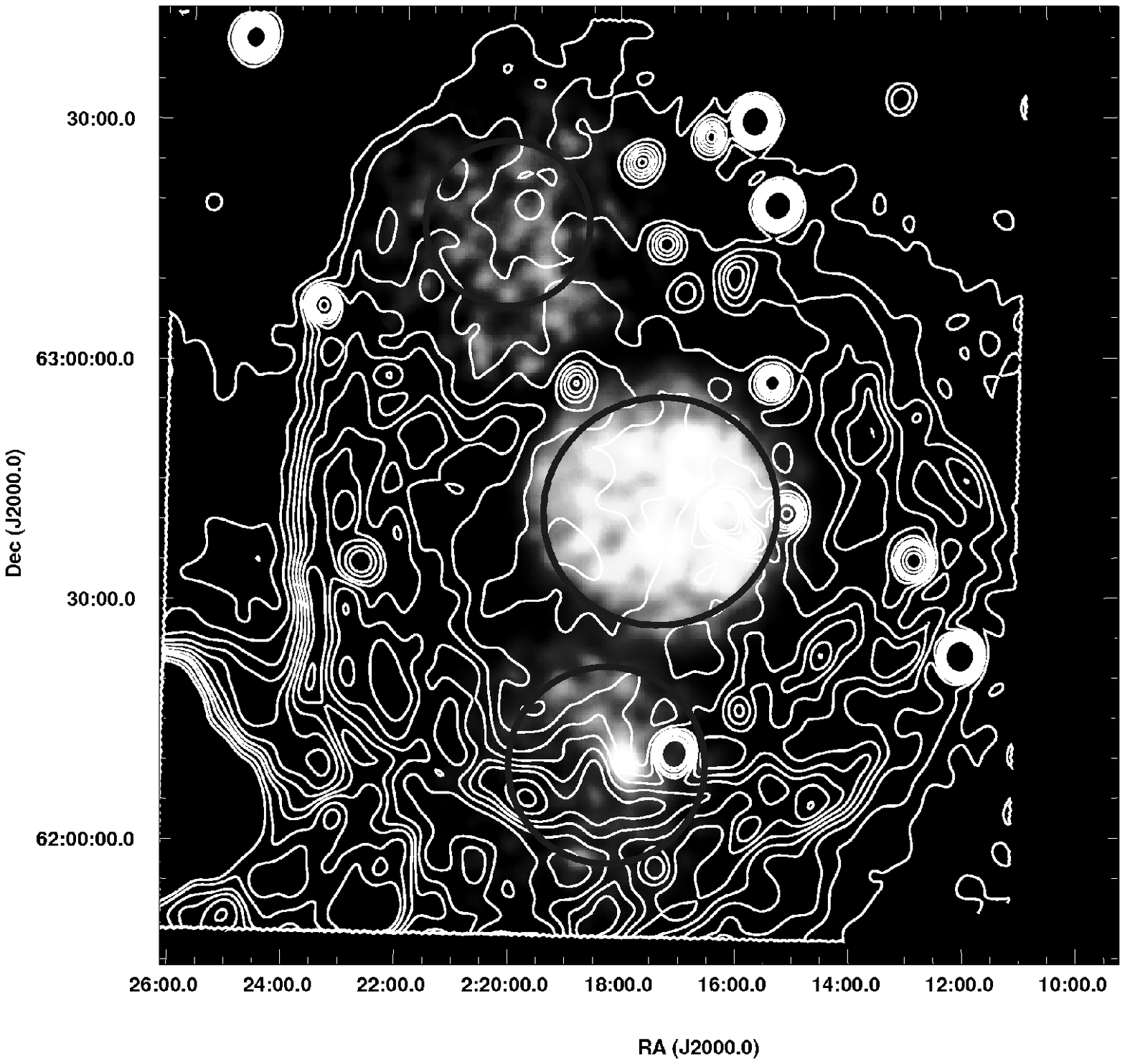}}
\caption{ (left) The broadband radio spectrum of HB3 based on
measured flux densities for this SNR as compiled by \citet{Tian05}
as fit with the two-component model (synchrotron and thermal
bremsstrahlung). (right) {\it ASCA} GIS mosaicked image of HB3:
the image has been background-subtracted, exposure-corrected and
smoothed with a Gaussian of 1 arcminute. The pixel range of the
X-ray emission is 0 to 1.43$\times$10$^{-4}$ counts sec$^{-1}$
pixel$^{-1}$. Radio emission (408 MHz) has been overlaid in white
contours: the contour levels are 60, 65, 70, 75, 80, 85, 90, 95,
100 110 and 120 mJy/beam. The regions of spectral extraction are
marked with circles and ellipses.\label{fig1}}

\end{figure}

\begin{deluxetable}{lcccccccc}
\tablecaption{Summary of {\it ASCA} GIS Observations of HB3
(G132.7$+$1.3)\label{Table1}} \tablewidth{0pt}
\tablehead{
& & & & & \colhead{GIS2} & \colhead{GIS2} & \colhead{GIS3} & \colhead{GIS3} \\
& & & & & \colhead{Effective} & \colhead{Count} &
\colhead{Effective} &
\colhead{Count}\\
& & \colhead{Sampled} & \colhead{Right} & & \colhead{Exposure} &
\colhead{Rate} & \colhead{Time} & \colhead{Rate}\\
\colhead{Sequence} & \colhead{Observation} & \colhead{Region} &
\colhead{Ascencion} & \colhead{Declination} & \colhead{Time} &
\colhead{(cts}
& \colhead{Exposure} & \colhead{(cts}\\
\colhead{Number} & \colhead{Date} & \colhead{of HB3} &
\colhead{(J2000.0)} & \colhead{(J2000.0)} & \colhead{(sec)} &
\colhead{sec$^{-1}$)} & \colhead{(sec)} & \colhead{sec$^{-1}$)}}
\startdata 54009000 & 27 August 1996 & South & 02 18 43.7 & $+$62
08 29 & 13406 &
0.2591 & 13404 & 0.2660 \\
54009010 & 28 August 1996 & Center & 02 17 07.5 & $+$62 42 55 &
29720 &
0.3806 & 31633 & 0.4075\\
54009020 & 29 August 1996 & North & 02 19 53.9 & $+$63 16 53 &
9261 & 0.2399 & 9260 & 0.2585
\enddata
\tablecomments{The units of Right Ascencion are hours, minutes and
seconds and the units of Declination are degrees, arcminutes and
arcseconds. Count rates are for the energy range 0.7--10.0 keV.}
\end{deluxetable}

\begin{deluxetable}{lcccc}
\tablecaption{Summary of Spectral Fits to {\it ASCA} GIS Spectra
of the Different Regions of HB3\tablenotemark{a}\label{Table2}}
\tablewidth{0pt}
\tablehead{
& & \colhead{PHABS$\times$(VAPEC} & \colhead{PHABS$\times$(VAPEC} \\
\colhead{Parameter} & \colhead{PHABS$\times$APEC} &
\colhead{+APEC+Gaussian)} & \colhead{+Power Law+Gaussian)} &
\colhead{PHABS$\times$APEC}} \startdata
Region & South & Center & Center & North \\
$\chi$$_{\nu}^2$ ($\chi$$^2$/DOF) & 1.15 (451.33/394) & 1.19
(458.26/386)
& 1.21 (467.82/386) & 1.22 (481.18/394) \\
$N$$_H$ (10$^{22}$ cm$^{-2}$) & 0.55$^{+0.17}_{-0.15}$ &
0.44$^{+0.11}_{-0.12}$
& 0.47$^{+0.11}_{-0.27}$ & 0.16$^{+0.08}_{-0.06}$ \\
$kT$$_{low}$ (keV) & 0.39$^{+0.15}_{-0.09}$ &
0.22$^{+0.06}_{-0.04}$
& 0.21$^{+0.11}_{-0.07}$ & 0.54$^{+0.05}_{-0.10}$\\
Mg & 1.0 (frozen) & 2.4$^{+1.1}_{-1.2}$ & 2.9($>$1.0) & 1.0 (frozen)\\
Si & 1.0 (frozen) & 4.2$^{+3.0}_{-2.2}$ & 4.6($>$2.0) & 1.0 (frozen) \\
EM$_{low}$\tablenotemark{b}/(4$\pi$$d$$^2$/10$^{-14}$) (cm$^{-5}$)
& 2$\times$10$^{-2}$ & 0.17 & 0.19 & 3$\times$10$^{-3}$ \\
$n$$_e$ (cm$^{-3}$) & 0.14$f$$^{-1/2}$ & 0.35$f$$^{-1/2}$ &
0.38$f$$^{-1/2}$
& 0.08$f$$^{-1/2}$ \\
$E$$_{gauss}$ (keV) & \nodata & 1.26 & 1.22 & \nodata \\
Normalization\tablenotemark{c} & \nodata & 4.1$\times$10$^{-4}$ &
2.0$\times$10$^{-4}$ & \nodata\\
$kT$$_{high}$ (keV) & \nodata & 2.2$^{+3.8}_{-0.2}$ & \nodata & \nodata \\
EM$_{high}$\tablenotemark{b}/(4$\pi$$d$$^2$/10$^{-14}$)
(cm$^{-5}$) & \nodata
& 1.3$\times$10$^{-3}$ & \nodata & \nodata \\
$\Gamma$ & \nodata & \nodata & 2.6$^{+1.8}_{-1.4}$ & \nodata \\
Normalization\tablenotemark{d} & \nodata & \nodata &
7.4$\times$10$^{-4}$ &
\nodata  \\
Absorbed Flux\tablenotemark{e} (ergs cm$^{-2}$ sec$^{-1}$) &
4.2$\times$10$^{-12}$ & 1.6$\times$10$^{-11}$ &
1.6$\times$10$^{-11}$
& 3.5$\times$10$^{-12}$ \\
Unabsorbed Flux\tablenotemark{e} (ergs cm$^{-2}$ sec$^{-1}$) &
2.4$\times$10$^{-11}$ & 6.8$\times$10$^{-11}$ &
7.0$\times$10$^{-11}$
& 5.9$\times$10$^{-12}$ \\
Luminosity\tablenotemark{e} (ergs sec$^{-1}$) &
1.1$\times$10$^{34}$ & 3.3$\times$10$^{34}$ & 3.4$\times$10$^{34}$
& 2.8$\times$10$^{33}$
\enddata
\tablenotetext{a}{All quoted errors are 90\% confidence
intervals.} \tablenotetext{b}{Emission measure ($\int n_e n_H$ dV
where $d$ is the distance in cm and $n$$_e$ and $n$$_H$ are the
electron and H densities in cm$^{-3}$).} \tablenotetext{c}{In
units of total photons/cm$^{2}$/sec in the line.}
\tablenotetext{d}{In units of photons/keV/cm$^{2}$/sec at 1 keV.}
\tablenotetext{e}{For the energy range 0.7--10.0 keV. The
luminosity estimates are for an assumed distance of 2 kpc to HB3.}

\end{deluxetable}

\end{document}